\documentstyle[epsfig]{article}
\def\func#1{\mathop{\rm #1{}}\nolimits}
\def\eps{\varepsilon}
\newcommand{\Partl}[2]{\mathop{\frac{\partial {#1}}{\partial {#2}}}\nolimits}

\newcommand{\gtsimscript}{\mathrel{\protect\raisebox{-0.5ex}{$\stackrel{\scriptstyle >}
    {\sim}$}}}
\begin{document}
\title{The Quantum-Classical Crossover in\\ the Adiabatic Response of Chaotic Systems}
\author{Ophir M. Auslaender\thanks{Corresponding author:
ophir@wis.weizmann.ac.il (present address: Dept. of Condensed
Matter Physics, Weizmann Institute of Science, Rehovot  76100,
Israel)} $~$and Shmuel Fishman \\ Department of Physics, Technion,
Haifa 32000, Israel}
\date{\today}
\maketitle
\begin{abstract}
The autocorrelation function of the force acting on a slow
classical system, resulting from interaction with a fast quantum
system is calculated following Berry-Robbins and Jarzynski within
the leading order correction to the adiabatic approximation. The
time integral of the autocorrelation function is proportional to
the rate of dissipation. The fast quantum system is assumed to be
chaotic in the classical limit for each configuration of the slow
system. An analytic formula is obtained for the finite time
integral of the correlation function,  in the framework of random
matrix theory (RMT), for a specific dependence on the
adiabatically varying parameter. Extension to a wider class of RMT
models is discussed. For the Gaussian unitary and symplectic
ensembles for long times the time integral of the correlation
function vanishes or falls off as a Gaussian with a characteristic
time that is proportional to the Heisenberg time, depending on the
details of the model. The fall off is inversely proportional to
time for the Gaussian orthogonal ensemble. The correlation
function is found to be dominated by the nearest neighbor level
spacings. It was calculated for a variety of nearest neighbor
level spacing distributions, including ones that do not originate
from RMT ensembles. The various approximate formulas obtained are
tested numerically in RMT. The results shed light on the quantum
to classical crossover for chaotic systems. The implications on
the possibility to experimentally observe deterministic friction
are discussed.
\end{abstract}

%%%%%%%%%%%%%%
%%%%%%%%%%%%%%
\section{Introduction}
%%%%%%%%%%%%%%
%%%%%%%%%%%%%%
Dissipation of energy from a physical system to a thermal bath
takes place as a result of a fluctuating force that acts on the
system because of its coupling to the bath. The dissipative
friction force is proportional to the correlation function of the
fluctuating force. The force resulting from coupling to a chaotic
system is also fluctuating. The question that will be studied in
the present paper is on what time scales it leads to friction and
what is the relation of this friction to the autocorrelation
function of the fluctuating force. An example is slow particle
coupled to a fast particle so that the motion of the fast particle
is chaotic for each position of the slow particle.

Various models for dissipation of energy from a slow particle by a
fast one have been developed. To our knowledge the first models of
this type were introduced in the context of nuclear physics
\cite{Hill52}. In particular, a model where particles move within
a region bounded by a deforming boundary, modeling the  nuclear
surface was studied \cite{Blocki78}. The energy transferred
between the boundary and particles enclosed inside was calculated
classically and quantum mechanically in the framework of some
approximations. Recently some detailed numerical simulations were
performed along these lines and the regime of validity of various
approximations was tested \cite{Burgio95}. The dissipation for a
wide class of model systems was explored by Wilkinson and Austin
\nocite{Wilkinson95,Wilkinson90,Austin92} [4-6] in the framework
of random matrix theory (RMT), relevant for a situation where the
motion of the fast particles is chaotic. The relation to
Landau-Zener tunneling was also studied in these works. A
different RMT model was studied by Mizutori and Aberg
\cite{aberg}. In all these studies dissipation was found. In
addition, a different approach, aimed at emphasizing the relation
to many body problems was introduced \cite{Bulgac98}. A unified
picture of many of these models has recently been presented by
Cohen \cite{CohenD98}. A systematic investigation of the
interaction of a slow system with a fast one is possible with the
help of multiple scale analysis. Under such conditions Ott
demonstrated \cite{Ott79} that the phase space volume enclosed by
the energy surface of the fast particle is an adiabatic invariant,
namely its change is much slower than that of the fast particle
Hamiltonian. It has been demonstrated for various conditions that
it is indeed an adiabatic invariant \cite{Brown87a}.
%%%%%%%%%%%%%%
%\subsection{A classical system}\label{SecClassic}
%%%%%%%%%%%%%%
In the present paper we study the behavior of a slow particle that
is coupled to a fast chaotic system. A model for such a system,
that is quite general, and has been studied by Berry and Robbins
(BR) \cite{Berry93} and by Jarzynski \cite{Jarzynski95} in the
framework of multiple scale analysis, is defined by the
Hamiltonian:
\begin{equation}\label{BRJhamiltonian}
{{\cal H}}=\frac{1}{2M}{{{\bf P}}}^2+{h}\left({\bf R},{\bf z}\right).
\end{equation}
The phase space coordinates of the slow particle are $\left({\bf
P},{\bf R}\right)$ and its mass is $M$. For simplicity it is
coupled only through its position to the fast system whose phase
space coordinates are ${\bf z}\equiv\left({\bf p},{\bf r}\right)$.
The latter system has the property that if ${\bf R}$ is kept fixed
it is fully chaotic. The crucial feature of the system we wish to
study in this work is that it exhibits a wide separation of time
scales- the evolution of the fast system, characterized by the
time scale $T_{\mbox{fast}}$, is so rapid that it explores all of
the phase space accessible to it energetically before the slow
particle, characterized by the time scale $T_{\mbox{slow}}$, moves
appreciably. The adiabaticity parameter is $\eps\sim
T_{\mbox{fast}}/ T_{\mbox{slow}}$. One way to realize this is to
couple two particles with a mass ratio of $m/M=\eps^2\ll 1$, as
one can see by rescaling the equations of motion. We now turn to
analyze the dynamics generated by the Hamiltonian
(\ref{BRJhamiltonian}) with the approximation that the slow
particle evolves under the influence of the {\em average} force
exerted on it by the fast system, which can be treated as a system
described by a slowly varying Hamiltonian. The time dependence of
this Hamiltonian is determined by the dynamics of the slow
particle. First the classical dynamics is outlined and later the
quantum mechanical behavior is summarized. In the case of
(\ref{BRJhamiltonian}) the average force is given by:
\begin{equation}\label{AvgForce}
{\bf F}(\tau_a)=-\int {\bf dz}~\rho\left({\bf z},\tau_a \right)
\partial_{\bf R}{h\left({\bf z},{\bf R}(\tau_a)\right)},
\end{equation}
where $\rho({\bf z},\tau_a)$ is a normalized probability density
in the fast particle phase space. A formalism that includes {\em
fluctuations} was developed by Jarzynski \cite{Jarzynski95}. The
results of this paper do not depend on these fluctuations and
therefore the formalism of BR will be used. The probability
density satisfies the Liouville equation:
\begin{equation}\label{BRliouville}
\eps\Partl{}{\tau_a}{\rho\left({\bf z},\tau_a\right)}= \Big\{
h\left({\bf z},{\bf R}(\tau_a)\right),\rho\left({\bf
z}\!,\tau_a\right) \Big\}_{\bf z},
\end{equation}
written in a way that emphasizes that the evolution of the fast
system is indeed on a much shorter time scale than the time scale
on which the fast Hamiltonian changes. $\{ \}_{\bf z}$ denotes
Poisson brackets taken with respect to ${\bf z}$. With the aid
of the multiple scale expansion:
\begin{equation}\label{epsExpansion}
\rho\left({\bf z},\tau_a\right)=\sum_{l=0}^\infty \eps^l \rho_l
\left({\bf z},\tau_a \right),
\end{equation}
Berry and Robbins \cite{Berry93} were able to calculate the force acting
on the slow particle up to first order in $\eps$:
\begin{equation}\label{BRforce}
{\bf F}\approx {\bf F}_0+\eps{\bf F}_1.
\end{equation}
To leading order, the force is given by the classical analogue of
the Born-Oppenheimer force:
\begin{equation}\label{F0}
{F_0}_i(\tau_a)=-\partial_{R_i}{E\left({\bf R}\right)},
\end{equation}
where $E\left({\bf R}\right)$ is chosen such that the phase space
volume enclosed by the energy surface of the fast particle,
$\Omega\left(E\left({\bf R}\right),{\bf R}\right)$, is constant.
The leading correction to ${\bf F}_0$ includes a velocity
dependent force:
\begin{equation}\label{F1} {F_1}_i(\tau_a)=-\sum_j
K_{ij}{\dot R}_j;~~~~~~~~K_{ij}\equiv
\Sigma^{-1}\partial_{E}\left[\Sigma(E,{\bf R}) \frac12
I_{ij}\left(E,{\bf R}\right)\right]_{E=E({\bf R})},
\end{equation}
where $\Sigma(E,{\bf R})\equiv\partial_{E}\Omega(E,{\bf R})$ and:
\begin{eqnarray}\label{BRIij} I_{ij}\left(E,{\bf R}\right)&=&2
\int_{0}^{\infty}\!\!\! dt'~C_{ij}\left(E,{\bf R};t'\right);\\
\label{BRCij} C_{ij}\left(E,{\bf
R};t'\right)&\equiv&\Bigg<\partial_{R_i}\tilde{h} \left({\bf
z}_{t'}({\bf z},{\bf R}),{\bf R}\right)
\partial_{R_j}\tilde{h}\left({\bf z},{\bf R}\right)\Bigg>_{E,{\bf
R}}.
\end{eqnarray}
In the last equation:
\begin{equation}\label{LiouvilleAverage}
\left<\ldots\right>_{E,{\bf R}}\equiv\int {\bf
dz}~\rho_0\left({\bf z},\tau_a \right) \ldots = \Sigma^{-1}(E,{\bf
R})\int {\bf dz}~\delta\left(E-h({\bf z},{\bf R})\right)\ldots
\end{equation}
denotes the microcanonical average and $\tilde{h}\left({\bf z},
{\bf R}\right)\equiv h\left({\bf z}, {\bf R}\right)-E({\bf R})$.
Finally, ${\bf z}_{t'}({\bf z},{\bf R})$ in (\ref{BRCij}) is the
classical trajectory obtained by integrating the equations of
motion generated by $h\left({\bf z}, {\bf R}\right)$ with fixed
${\bf R}$ backwards from ${\bf z}$ for the time $t'$. In addition
to the velocity dependent force (\ref{F1}), and to the same order
in $\eps$, there is a force that does not depend on velocity
\cite{Jarzynski93b}. This force can be expressed as a gradient of
a time dependent potential, and therefore is a correction to the
Born-Oppenheimer force (\ref{F0}). Because of the form of ${\bf
F}_1(\tau_a)$, it describes two qualitatively different forces.
The first of these is geometric magnetism, and it is related to
the antisymmetric part of $K_{ij}$. This force has been studied
analytically in the systems under discussion by BR \cite{Berry93}
and numerically by Berry and Sinclair \cite{Berry97a}. The second
force is associated with the symmetric part of $K_{ij}$ and is
related to deterministic friction. This force has been studied in
\cite{Wilkinson90,Jarzynski92,Berry93,Jarzynski95}. A central
question that can be addressed at this point is under which
conditions does the slow particle feel friction due to the
velocity dependent force ${\bf F}_1$. In order for this to happen
$K_{ij}$ of (\ref{F1}) has to have a positive definite symmetric
part.

The behavior when the fast system is quantum mechanical, to first
order in $\varepsilon$, has also been studied by BR
\cite{Berry93}. They found that in this case $K_{ij}$ is an
antisymmetric tensor, meaning that the system exhibits {\em only}
geometric magnetism and {\em no} friction. This difference is a
result of the discreteness of the quantum spectrum. The quantum
correlation function corresponding to the symmetric part of
$C_{ij}$ of (\ref{BRCij}) is:
\begin{eqnarray}\label{BRCijQ}
\hspace{-2cm}\frac12\Big\{C_{ij}(n;t)+C_{ji}(n;t)\Big\}&=&\nonumber\\
&&\hspace{-4cm}=\sum_{m\neq
n}\Re\left(\left<n\left|\partial_{R_i}{\widehat
h}\right|m\right>\left<m\left|\partial_{R_j}{\widehat
h}\right|n\right>
\right)\cos\left[\frac{t}{\hbar}\left(E_n-E_m\right)\right],
\end{eqnarray}
and the infinite time integral over it vanishes \cite{Berry93}. In
order to understand how the crossover between the classical and
quantum behavior occurs, it is instructive to calculate the
integral of the correlation function over a finite time. Since the
discordance between the quantum and the classical models appears
in the symmetric part of $K_{ij}$ it can be studied for the case
where ${\bf R}$ is replaced be a scalar, time dependent parameter,
$X$. In this case $\frac12\Big\{C_{ij}(n;t)+C_{ji}(n;t)\Big\}$
will be denoted by $C(t)$ for simplicity ($R_i$ and $R_j$ on the
RHS of (\ref{BRCijQ}) will then be replaced by the scalar
parameter $X$). Following BR we assume in the calculation that the
initial state is an eigenstate of the Hamiltonian,
$\left|n\right>$. It was verified by BR that their result holds
also if the initial state is a mixture. Our calculation can also
be extended to a mixture leading to the same results. If the
initial state is a pure state but not an eigenstate of the
Hamiltonian one can check that within the assumptions of the paper
the results are similar to the ones found if the initial state is
an eigenstate of the Hamiltonian. The finite time integral that
should be calculated then is:
\begin{equation}\label{InDef}
I(t)=\int_0^t C(t')dt'.
\end{equation}
The correlation function and its integral may depend on the
initial state $n$. This dependence has been suppressed in the
notation for simplicity. In what follows, it will become clear
that it is not important for the results of the present paper.
Taking the classical limit $\hbar\rightarrow 0$ for any finite $t$
and then the limit $t\rightarrow\infty$ should result in a
non-vanishing value of $I(\infty)$, while for any finite value of
$\hbar$, $I(\infty)$ should vanish. The friction on the time scale
$t$ is proportional to $I(t)$ as can easily be inferred from
(\ref{F1}) and (\ref{BRIij}). The experimental meaning of this
statement will be clarified in what follows. In order to
understand the mechanism of this discordance, BR studied a model
correlation function where the levels were equally spaced. They
found that the function is periodic in time with period
$t_p=\hbar/\Delta E$, where $\Delta E$ is the level spacing.
Moreover, in the classical limit, which in their model corresponds
to taking $t_p\rightarrow\infty$, $C(t)$ approaches the classical
correlation function. For systems whose classical dynamics is
chaotic, the energy levels are not equally spaced, but rather are
distributed according to RMT \cite{Bohigas84}. The long time
behavior of $I(t)$ is determined by the levels nearest to $n$,
namely $n\pm1$, as can be seen from (\ref{BRCijQ}). Therefore, one
may expect behavior different from the one found for the equally
spaced spectrum. The natural question to ask is whether there is a
characteristic time scale for the crossover between the quantum
behavior of the integral $I(t)$ and its classical behavior. The
most na\"{\i}ve answer to this question is that the characteristic
time scale is the Heisenberg time, because it is the only time
scale in the problem, and it is on this time scale that the
quantum to classical crossover usually takes place. On the other
hand, one can argue that there is no time scale for this crossover
at all \cite{izrailev}. In RMT the probability for two consecutive
levels to be separated by a distance $s$ behaves like $s^\beta$
for small spacings \nocite{Wigner56,Landau55,Porter65,Mehta91}
[19-22]. Consequently, $\left<I_\beta(t)\right>\sim t^{-\beta}$
for long times, where here $\left<\ldots\right>$ denotes the RMT
ensemble average, and $I_\beta$ is the integral (\ref{InDef}) for
some $\beta$. The answer given by the analysis presented in this
paper is surprising. For the Gaussian orthogonal ensemble (GOE)
($\beta=1$) one indeed finds that $\left<I_\beta(t)\right>$ decays
like $1/t$, but for the Gaussian unitary ensemble (GUE)
($\beta=2$) one finds that it decays like a Gaussian with a
characteristic time proportional to the Heisenberg time or
vanishes after the Heisenberg time depending of the parametric
dependence on $X$. The integral $I_{\beta}(t)$ was also calculated
for other values of $\beta$. Why is the nature of the decay of
$I_\beta(t)$ important? There is the quantum-classical discordance
that has already been mentioned, and one would like to analyze the
scale that is required to observe the crossover between the
regimes. It is relevant for some experiments, that will be
mentioned below. In addition, there is the issue of the relevance
of such an effect. The time over which the correlation function
decays should be compared with other time scales present in the
specific system studied. One such time scale is
$T_2\sim\eps^{-2}$, which is the time scale for the breakdown of
the first order of the multiple scale analysis. Non-perturbative
effects, such as Landau-Zener tunneling, become important on a
time scale of $T_{LZ}$. In realistic experiments there is also the
time scale for quantum decoherence $T_\phi$. In order to observe
the classical to quantum crossover discussed in the present work
$\left<I_\beta(t)\right>$ should exhibit substantial decay for $t
\ll \min\left(T_2,T_{LZ}\right)$  and of the order of  $T_\phi$.

The model discussed in the present work is relevant for some
experimental situations. Consider for example a molecular beam
prepared in a classical configuration, where initially many levels
are substantially populated. The beam travels in a slowly varying
field \cite{DCPC}. Consequently the internal dynamics in the
molecules is in a slowly varying potential. On short time scales
the behavior is classical, the integral of the correlation
function is positive and energy is absorbed in the motion of the
internal degrees of freedom. On longer time scales the integral of
the correlation function decays to zero and therefore one realizes
that actually no energy is absorbed by the molecules. The outcome
of the experiment depends on the time scale of decoherence,
$T_\phi$, that is, the energy absorption by the internal degrees
of freedom is proportional to $I_\beta(T_\phi)$. Another example
is of quantum dots where parameters are varied adiabatically, like
in pumping experiments, but with dots that are closed, so that
their spectrum is discrete \cite{marcus}.

In Section \ref{SecRMT} a specific RMT model is defined. For this
model the ensemble average of the integral of the correlation
function (\ref{InDef}) is calculated analytically. It is
demonstrated that most of the contribution for long times
originates from the nearest neighbor levels. In Section \ref{NND}
the integral of the correlation function (\ref{InDef}), predicted
by the nearest neighbor level spacing distribution, is calculated
for various distributions, including some that are not related to
RMT models. In Section \ref{SecDiscussion} the results of this
work are analyzed and discussed.

%%%%%%%%%%%%%%%%%
%%%%%%%%%%%%%%%%%
\section{Random matrix models}\label{SecRMT}
%%%%%%%%%%%%%%%%%
The main purpose of this paper is to study (\ref{BRCijQ}) and its
finite time integral (\ref{InDef}) in the framework of RMT. The
reason for this is that random matrices describe many
characteristic properties of realistic quantum-chaotic systems
\nocite{Bohigas84,Andreev95,Bogomolny96,Zirnbauer99,Feingold90}
[17, 25-28]. For simplicity our random matrices will depend on one
external parameter $X$, so that we shall study the one dimensional
version of (\ref{BRCijQ}), where the vector of parameters $({\bf
R})$ is replaced by the scalar $X$. For each random matrix we
shall be able to calculate both $C(t)$ and $I(t)$.

%%%%%%%%%%%%%%%%%%%%%%%%%%%%%%%%%%%%%%%%
\subsection{A simple RMT model}
%%%%%%%%%%%%%%%%%%%%%%%%%%%%%%%%%%%%%%%%
We wish to construct a random matrix model for some of the levels
of a system whose quantum Hamiltonian depends on some parameter.
The $N$ levels we wish to simulate by the random matrix lie within
an energy strip of width $\delta E(N)$, that depends on $N$. Later
on we shall be interested in studying the semiclassical limit. The
meaning of taking this limit in the present context is to increase
the density of levels in the $\delta E$-strip: in the classical
limit the spectrum becomes continuous. We shall  work with the
well studied Gaussian Ensembles \cite{Porter65,Mehta91}. These are
defined through four parameters: $\beta$ which defines the
symmetry of the random matrices, their dimension $N$, the mean
value of their elements and their variance (given through the
parameter $\mu^2$). All of these need to be chosen carefully in
terms of parameters of the physical system, being simulated by the
random matrix. The symmetry of the ensemble should be chosen to
correspond to the real system. If the latter exhibits time
reversal symmetry then the ensemble is the orthogonal one
$(\beta=1)$. If the system does not exhibit this symmetry then the
ensemble is unitary $(\beta=2)$. The mean value of the matrix
elements can be chosen to be zero, which corresponds to setting
the ensemble average of the reference level $\left<E_n\right>=0$.
The mean level density satisfies the semi-circle law
\cite{Wigner55,Porter65,Mehta91}:
\begin{eqnarray}\label{semicircle}
\overline{\rho}_x(x)\approx
\left\{\begin{array}{cc}
\frac{2N}{\pi}\sqrt{1-x^2} &~~|x|<1\\
                        0  &~~\mbox{otherwise}\\
        \end{array}\right.
\end{eqnarray}
for large $N$, where we have used the definition: $x\equiv
E/\sqrt{4\beta\mu^2N}$. In subsection \ref{SecShort} the relation
between the parameters of the RMT model and the ones of the
physical system will be discussed.

\begin{figure}
\centering\epsfig{file=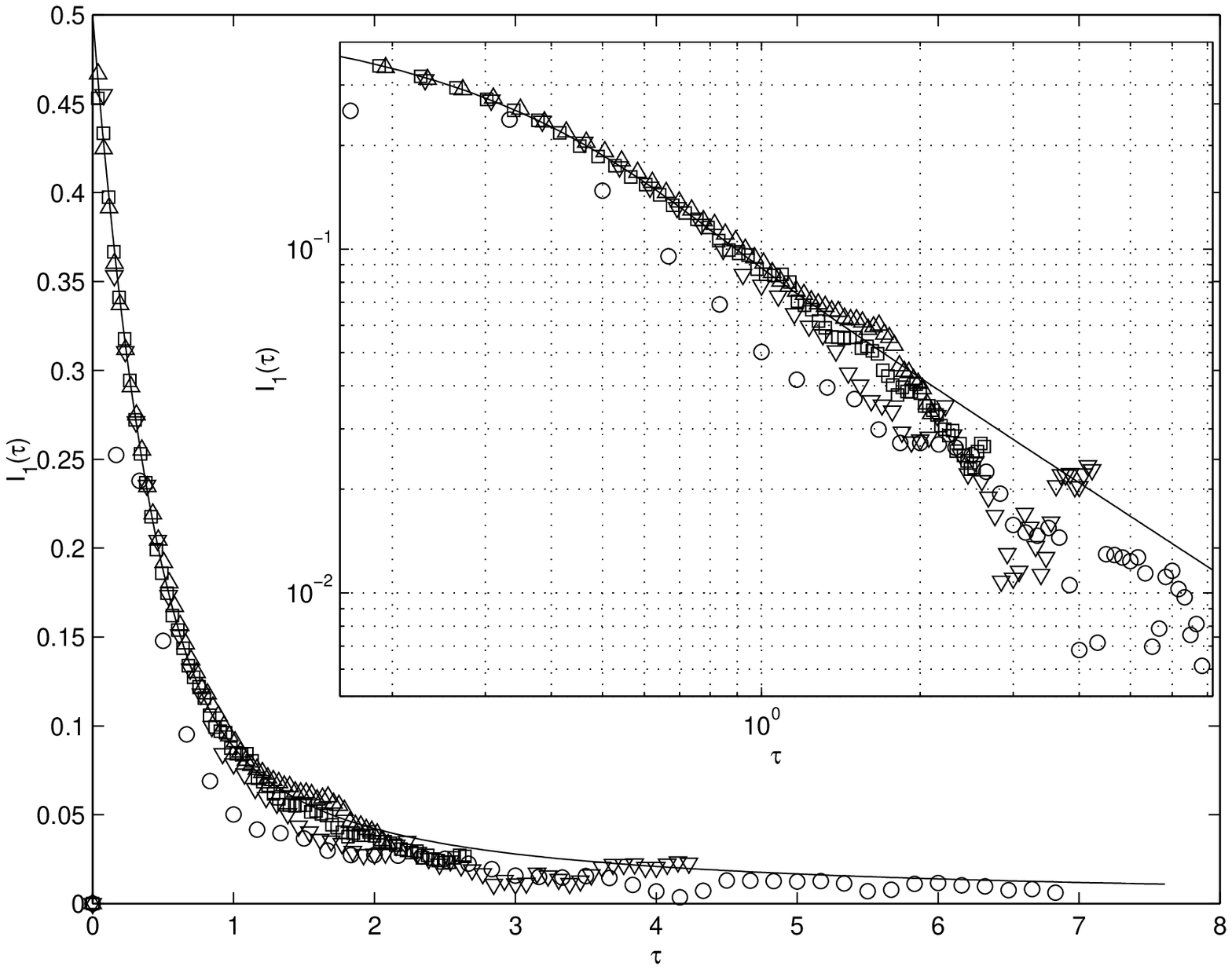,height=10cm,angle=90}
\caption{The integral of the correlation function for GOE.
Numerical results for $N=3$ ($\scriptscriptstyle\bigcirc$), $N=13$
($\scriptstyle\bigtriangledown$), $N=53$ ($\scriptstyle\Box$) and
$N=103$ ($\scriptstyle\bigtriangleup$) are shown. Also shown is
the large N approximation (Eq.~\protect\ref{IGOE}) (line). The
inset shows the long time behavior on a log-log scale. The number
of ensemble members used is $10^4$ and the errors are of the order
of $\Delta I_1\approx0.01$.} \label{figGOE}
\end{figure}

\begin{figure}
\centering\epsfig{file=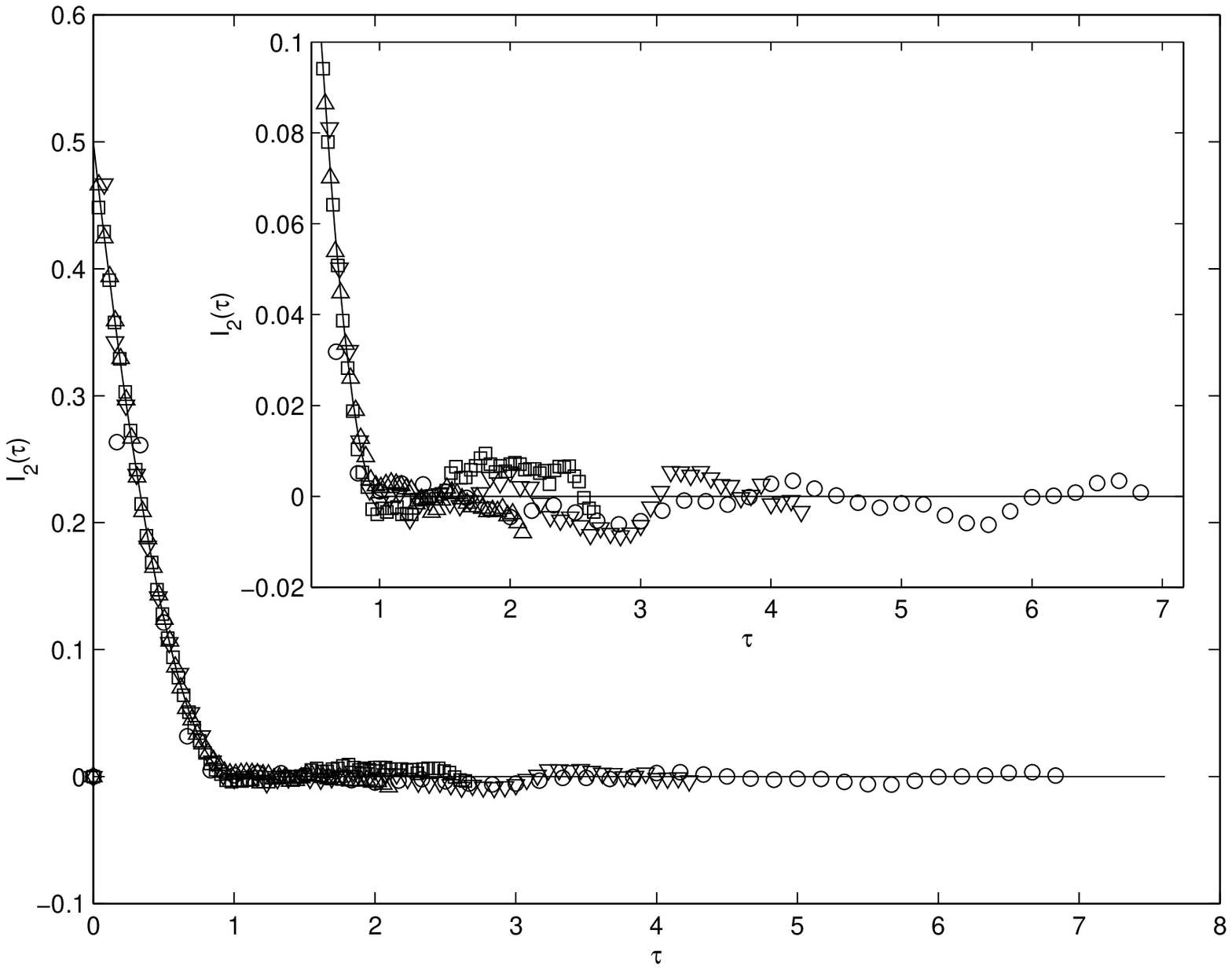,height=10cm,angle=90}
\caption{The integral of the correlation function for GUE.
Numerical results for $N=3$ ($\scriptscriptstyle\bigcirc$), $N=13$
($\scriptstyle\bigtriangledown$), $N=53$ ($\scriptstyle\Box$) and
$N=103$ ($\scriptstyle\bigtriangleup$) are shown. Also shown is
the large N approximation (Eq.~\protect\ref{IGUE}) (line). The
inset emphasizes the long time behavior. The number of ensemble
members used is $10^4$ and the errors are of the order of $\Delta
I_2\approx0.01$.} \label{figGUE}
\end{figure}

We shall use the Hamiltonian introduced by Austin and Wilkinson
\cite{Austin92} and model a parameter dependent system by the
$N\times N$ random matrix:
\begin{equation}\label{model}
H(X)=H_1 \cos{X}+H_2\sin{X},
\end{equation}
where $H_{1,2}$ are $N\times N$ random matrices from the same GOE
or GUE ensemble. There are three advantages to working with $H(X)$: (a) it
belongs to the same ensemble that $H_{1,2}$ belong to, (b) the
derivatives of its matrix elements belong to the same ensemble too
because:
\begin{equation}\label{modelDer}
dH(X)/dX=-H_1 \sin{X}+H_2\cos{X},
\end{equation}
(c) the matrices $H(X)$ and $dH(X)/dX$ are statistically independent.
If we insert $H(X)$ and $dH(X)/dX$ into Eq.~\ref{BRCijQ}, and then
perform the ensemble average, we obtain:
\begin{equation}\label{CnDef}
C_\beta(t)=\Bigg<\sum_{m\neq
n}\left|\left(dH(X)/dX\right)_{n,m}\right|^2\cos\left[\frac{t}{\hbar}
\left(E_n-E_m\right)\right]\Bigg>,
\end{equation}
where $\left(dH(X)/dX\right)_{n,m} \equiv \left<n\left|d{\widehat
H}(X)/dX\right|m\right>$ and $\left<\ldots\right>$ denotes RMT
ensemble averaging. The subscript $\beta$ will denote the
symmetry: $\beta=1$ for GOE and $\beta=2$ for GUE. The correlation
function $C_\beta$ and its finite time integral $I_\beta$ are
ensemble averaged. The $\left<\ldots\right>$ will be dropped from
these quantities for notational simplicity.

The statistical independence of $dH(X)/dX$ and $H(X)$ implies
\begin{equation}\label{CnInd}
C_\beta(t)=\sum_{m\neq n}
\Bigg<\left|\left(dH(X)/dX\right)_{n,m}\right|^2 \Bigg>
 \Bigg<\cos\left[\frac{t}{\hbar}
\left(E_n-E_m\right)\right]\Bigg>,
\end{equation}
while the fact that $dH(X)/dX$ belongs to the same ensemble as $H(X)$ implies
\begin{equation}\label{ElementsAvg}
\Bigg<\left|\left(dH(X)/dX\right)_{n,m}\right|^2 \Bigg> =
\Bigg<\left|\left(H(X)\right)_{n,m}\right|^2 \Bigg> = \beta \mu^2
\end{equation}
for $m\ne n$, leading to:
\begin{equation}\label{CnNoCorr}
C_\beta(t)=\beta\mu^2\sum_{m\neq n}
\Bigg<\cos\left[\frac{t}{\hbar}\left(E_n-E_m\right)\right]\Bigg>.
\end{equation}

We would like to make the connection between
$C_\beta(t)/\beta\mu^2$ and the form factor:
\begin{equation}\label{formfactor}
K(t)=\int\left[\frac{1}{\overline{\rho}^2(E)}
\Big<\rho(E+\epsilon/2\overline{\rho}) \rho(E-\epsilon/2\overline{\rho})\Big>-
1\right]e^{i2\pi\epsilon \tau}d\epsilon,
\end{equation}
where $\rho(E)=\sum_i \delta(E_i-E)$ is the density of states and
$\overline{\rho}(E)$ is the smoothed density of states. The
variable $\epsilon$ is the energy measured in units of the mean
level spacing $1/\overline{\rho}(E)$ and $\tau=t/T_H$ is time in
units of the Heisenberg time, $T_H=h\overline{\rho}(E)$.

Eq.~\ref{CnNoCorr} can be written in the following form:
\begin{equation}
C_\beta(\tau)/\beta\mu^2=
\int\left[\frac{1}{\overline{\rho}^2(E)}
\Big<\rho(E+\epsilon/2\overline{\rho})\rho(E-\epsilon/2\overline{\rho})\Big>
-\delta(\epsilon) \right]e^{i2\pi\eps\tau}d\epsilon,
\end{equation}
where the $\delta(\epsilon)$ results from the omission of the term
$m=n$ in the sum (\ref{CnNoCorr}). Comparing the last equation
with (\ref{formfactor}) one can see that:
\begin{equation}
C_\beta(\tau)/\beta\mu^2=K(\tau)+\delta(\tau)-1.
\end{equation}
In this work we are mainly interested in the time integral of the
correlation function (\ref{InDef}):
\begin{equation}\label{IandK}
  I_\beta(\tau)/\beta\mu^2T_H=\int_0^\tau d \tau'
C_\beta(\tau')/\beta\mu^2=\left[1/2-\int_0^{\tau}d\tau'
\Big(1-K(\tau')\Big)\right].
\end{equation}
In the limit $\tau \rightarrow \infty$ the term in the square
brackets is just $R_2(\epsilon=0)$, the two point spectral
correlation function at zero energy separation. It vanishes as a
result of level repulsion.

In order to perform actual calculations we make use of the well
known form factor for GOE and GUE \cite{Mehta91}. It is standard
to define
\begin{equation}
  b(\tau)=1-K(\tau).
\end{equation}
For GOE it is given for example in Mehta's book (see
\cite{Mehta91} p. 137):
\begin{equation}\label{bGOE}
  b(\tau)=\left\{
  \begin{array}{lr}
    1-2\tau+\tau\ln\left[1+2\tau\right] & \tau\leq1 \\
    -1+\tau\ln\left[\frac{2\tau+1}{2\tau-1}\right] & \tau\geq1
  \end{array}
  \right.,
\end{equation}
from which one obtains:
\begin{equation}\label{IGOE}
  \frac{I_1(\tau )}{\mu^2T_H}=\left\{
  \begin{array}{lr}
\frac12-\left[\frac54\left(\tau-\tau^2\right)+\frac12
\left(\tau^2-\frac14\right)\ln\left[1+2\tau\right]\right] &
\tau\leq1 \\
\frac12-\left[\frac12\left(1-\tau\right)+\frac12\left(\tau^2-\frac14\right)\ln
\left[\frac{2\tau+1}{2\tau-1}\right]\right]& \tau\geq1
  \end{array}
  \right..
\end{equation}
For $\tau\rightarrow\infty$ $I_1(\tau )$ falls off asymptotically
as
\begin{equation} \label{IGOEa}
\frac{I_1(\tau)}{\mu^2T_H} \sim \frac{1}{12 \tau}.
\end{equation}

For GUE (see \cite{Mehta91} p. 95):
\begin{equation}\label{bGUE}
  b(\tau)=\left\{
  \begin{array}{cr}
    1-\tau & \tau\leq1 \\
    0      & \tau\geq1
  \end{array}
  \right.,
\end{equation}
from which one obtains:
\begin{equation}\label{IGUE}
  \frac{I_2(\tau )}{2\mu^2T_H}=\left\{
  \begin{array}{cr}
    \frac12-\left[\tau-\frac{\tau^2}2\right] & \tau\leq1 \\
    0 & \tau\geq1
  \end{array}
  \right..
\end{equation}

In order to compare the analytical results that hold in the
infinite $N$ limit with results for finite $N$, ensembles of the
$N\times N$ matrices $H_1$ and $H_2$ of (\ref{model}), belonging
to GOE or GUE were generated numerically. The integral of the
correlation function $I_\beta$ was then calculated numerically, by
ensemble averaging. The results are presented in
Figs.~\ref{figGOE}~\&~\ref{figGUE} for GOE and GUE respectively
and compared with (\ref{IGOE}) and (\ref{IGUE}). Units where
$\beta\mu^2T_H=1$ were used (see subsection \ref{SecShort}). The
numerical errors in the figures were calculated according to
\begin{eqnarray*} \Delta I_\beta=
\frac{1}{N_{ens}}\sqrt{\left<I_\beta^2\right>-\left<I_\beta\right>^2},
\end{eqnarray*}
where $N_{ens}$ is the number of matrices used in the ensemble average.

For completeness the results for GSE are obtained with the help of
(see \cite{Mehta91} p. 166):
\begin{equation}
  b(\tau)=\left\{
  \begin{array}{cr}
    1-\frac{1}{2}\tau+ \frac{1}{4}\tau\ln\mid 1-\tau\mid & \tau\leq 2 \\
    0                                                        & \tau\geq 2
  \end{array}
  \right..
\end{equation}
Following the calculation performed for the other ensembles one finds:
\begin{equation}\label{IGSE}
  \frac{I_4(\tau )}{4\mu^2T_H}=\left\{
  \begin{array}{cr}
\frac12- \left[\frac{1}{16}\left(14\tau-5\tau^2\right)
+\frac18\left(\tau^2-1\right)\ln\left|1-\tau\right|\right]
    &  \tau\leq 2 \\
0   & \tau\geq 2
  \end{array}
  \right..
\end{equation}

%%%%%%%%%%%%%%%%%%%%%%%
\subsection{Nearest neighbor spacing dominance and the long time limit}
%%%%%%%%%%%%%%%%%%%%%%%
\begin{figure}
\centering\epsfig{file=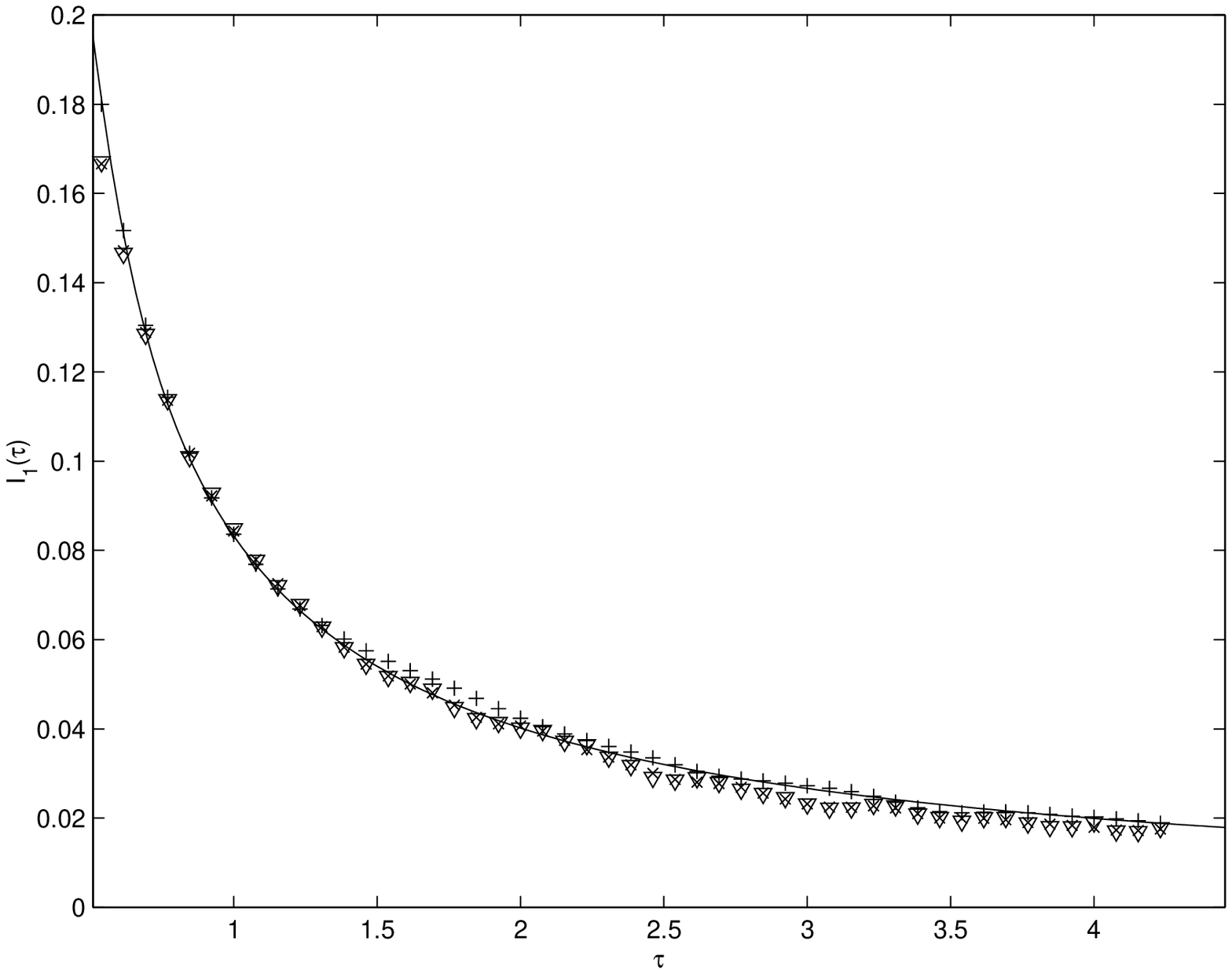,height=10cm,angle=90}
\caption{Testing the nearest neighbor approximation
(\protect\ref{CnApproxLong0}) for GOE. Full numerical results for
$N=13$, including all level spacings
($\scriptstyle\bigtriangledown$), using only nearest and second
nearest neighbors ($\scriptstyle\times$) and using only nearest
neighbors ($\scriptstyle+$). Also shown is the long time
approximation, where only nearest neighbor spacings are taken into
account (Eq.~\protect\ref{InApproxLong1GOE}) (line). The number of
ensemble members used was $10^5$ and the errors are of the order
of $\Delta I_1\approx0.005$.} \label{figGOEbruno}
\end{figure}

\begin{figure}
\centering\epsfig{file=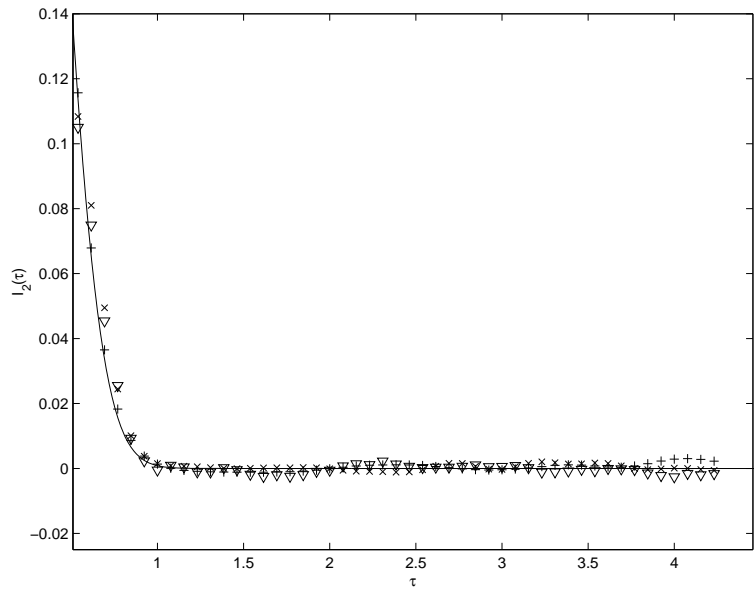,height=10cm,angle=90}
\caption{Same as Fig.~\protect\ref{figGOEbruno}, but for GUE,
compared with the long time approximation
(\protect\ref{InApproxLong1GUE}) (line). The number of ensemble
members used was $10^5$ and the errors are of the order of $\Delta
I_2\approx0.003$.} \label{figGUEbruno}
\end{figure}

The model (\ref{model}) is very specific in its dependence on the
parameter $X$. An important property of this model is the
statistical independence between $H(X)$ and $dH(X)/dX$. Such
independence holds to a good approximation for disordered systems
\cite{lerner}. It is reasonable to make this approximation also
for RMT models of chaotic systems. The reason is that most
eigenstates look random, are statistically independent of the
eigenvalues and therefore for many types of parametric
dependencies the matrix elements of $dH(X)/dX$ will look random
and independent of the spectrum. Although this argument is
reasonable for many types of parametric dependencies it is clearly
not general. For the asymptotic behavior much less is required,
since the long time asymptotics is dominated by the nearest
neighboring levels.  The reason for this dominance is that if
$\tau\gg 1$ the terms in the sum (\ref{CnDef}) oscillate wildly as
a function of $m$, so that the important net contribution is from
the terms nearest to being stationary. These are obviously
$m=n\pm1$. The approximation is therefore:
\begin{equation}\label{CnApproxLong0}
C_\beta(t)\approx2\Bigg<\left|\left(dH(X)/dX\right)_{n,n-1}\right|^2
\cos\left[\frac{t}{\hbar}\left(E_n-E_{n-1}\right)\right]\Bigg>
~~~~\mbox{for}~t\gg T_H.
\end{equation}
Only $m=n-1$ is required since we know that the matrix elements
and eigenvalue distributions are symmetric with respect to
reflection around the middle eigenvalue $n$. We further
approximate $\left|\left(dH(X)/dX\right)_{n,n-1}\right|^2$ by its
mean value and ignore the contribution from its fluctuations
resulting in:
\begin{equation}\label{CnApproxLong1}
C_\beta(t)\approx2\beta\mu^2\Bigg<\cos\left[\frac{t}{\hbar}
\left(E_n-E_{n-1}\right)\right]\Bigg>.
\end{equation}
Hence we ignored
\begin{equation}\label{DCn}
\Delta
C_\beta(t)=2\Bigg<\left(\left|\left(dH(X)/dX\right)_{n,n-1}\right|^2-\beta\mu^2
\right)
\cos\left[\frac{t}{\hbar}\left(E_n-E_{n-1}\right)\right]\Bigg>
\end{equation}
in the $t\gg T_H$ limit. In the framework of RMT
(\ref{CnApproxLong1}) takes the form:
\begin{equation}\label{CnApproxLong2}
C_\beta(\tau)/\beta\mu^2 \approx 2 \int_0^\infty \!\!\!\!ds~
P_\beta(s)\cos{(2 \pi \tau s)},
\end{equation}
where $s$ is the nearest neighbor spacing in units of the mean
level spacing $\Delta E=1/\overline{\rho}(0)$, the time $\tau$ is
measured in units of the Heisenberg time $T_H$ and $P_\beta(s)$ is
the probability distribution of $s$. The integral of the
correlation function is:
\begin{equation}\label{InApproxLong}
\frac{I_\beta(\tau)}{\beta\mu^2T_H} \approx 2 \int_0^\tau
d\tau'\int_0^\infty\!\!\!\!ds~ P_\beta(s) \cos (2 \pi\tau'
s)=\frac{1}{\pi}\int_0^\infty \!\!\!\!ds~
\frac{P_\beta(s)}{s}\sin(2 \pi\tau s).
\end{equation}
For the nearest neighbor level spacing distribution we use the
Wigner surmise \cite{Wigner56}, that is exact for $2\times2$
matrices, and takes the form (see Eq.~202 in \cite{Porter65}):
\begin{equation}\label{PsGOE}
P_1(s)=\frac\pi2 s \exp{\left[-\frac\pi4 s^2\right]}
\end{equation}
for GOE, and
\begin{equation}\label{PsGUE}
P_2(s)=\frac{32}{\pi^2} s^2 \exp{\left[-\frac4\pi s^2\right]}
\end{equation}
for GUE. The integral (\ref{InApproxLong}) can be calculated for
these distributions. For GOE one finds:
\begin{equation}\label{InApproxLong1GOE}
\frac{I_{1}(\tau)}{\beta\mu^2T_H}=\frac{1}{2}\exp{\left[-4 \pi
\tau^2 \right]} \func{erf}{\left[i~2 \sqrt{\pi} \tau\right]}/i,
\end{equation}
while for GUE one finds:
\begin{equation}\label{InApproxLong1GUE}
\frac{I_{2}(\tau)}{\beta\mu^2T_H}=2\tau
\exp{\left[-\frac{\pi^3}{4}\tau^2\right]}.
\end{equation}

A crucial approximation in the long time regime is
(\ref{CnApproxLong0}), where only the contribution of the nearest
neighbors is taken into account. This approximation is tested in
Figs.~\ref{figGOEbruno}~\&~\ref{figGUEbruno} for the model
(\ref{model}).  In the numerical test the nearest and
next-nearest neighbor spacings are taken from the center of the
matrix.  We see from these figures that the approximation
(\ref{CnApproxLong0}) is quite reasonable and it improves as time
increases. The reason is that as time grows the oscillations of
the various terms in (\ref{CnDef}) with energy become stronger,
enhancing the dominance of the nearest neighbor contributions. The
approximation is better for GOE than for GUE because the small $s$
weight in the integral (\ref{InApproxLong}) is larger.

The asymptotic behavior of $\func{erf}{\left[iy\right]}/i$ for
large real $y$ is $\exp{\!\left[y^2\right]}/(\sqrt{\pi} y)$
\cite{AS}. Therefore, we find for GOE:
\begin{equation}\label{InApproxLong2GOE}
\frac{I_{1}(\tau)}{\mu^2T_H} \sim \frac{1}{4 \pi
\tau}~~~~\mbox{for}~\tau\gg1,
\end{equation}
and there is {\em no} characteristic time for the crossover from
classical to quantum behavior. This decay is extremely close to
(\ref{IGOEa}) hence for large $\tau$ the contribution of the
nearest neighbor spacings accounts for nearly all of $I_1(\tau)$.

For GUE the falloff is much faster since in the large $\tau$ limit
the Gaussian in (\ref{InApproxLong1GUE}) dominates, and there {\em
is} a characteristic time $2/\pi^{3/2}$ (in units of the
Heisenberg time), for the crossover to quantum behavior. For long
time $I_2(\tau)$ is extremely small, while it is exactly zero
according to (\ref{IGUE}). Therefore also for GUE
(\ref{CnApproxLong0}) is an excellent approximation for the long
time limit.

For completeness let us present the results for the Gaussian
symplectic ensemble (GSE) (for which $\beta=4$), analogous to
(\ref{InApproxLong1GOE}) and (\ref{InApproxLong1GUE}). For this we
need to substitute the appropriate $P_\beta(s)$ into
(\ref{InApproxLong}). According to  Eq.~202 in \cite{Porter65}:
\begin{equation}\label{PsGSE}
P_4(s)=\frac{2^{18}}{3^6\pi^3} s^4
\exp{\left[-\frac{64}{9\pi}s^2\right]},
\end{equation}
leading to:
\begin{equation}\label{InApproxLongGSE}
\frac{I_4(\tau)}{\beta\mu^2T_H}=2
\left(1-\frac{3\pi^3}{32}\tau^2\right)\tau
\exp{\left[-\frac{9\pi^3}{64} \tau^2\right]}.
\end{equation}
One can immediately see that there is a characteristic time scale
for the quantum to classical crossover, as there was in the GUE
case. As for the GUE case $I_4(\tau)$ is extremely small for long
times, while (\ref{IGSE}) gives zero.

In Figs.~\ref{figGOEfluc}~\&~\ref{figGUEfluc} it is demonstrated
that the long time results of the simulation tend to the
analytical formulas as more and more members are included in the
ensemble.

%%%%%%%%%%%%%%%%%%%%%%%%%%%%%%%%%%%%%%%%%%%%%%
\subsection{The short time limit
and determination of the parameters of the RMT model} \label{SecShort}
%%%%%%%%%%%%%%%%%%%%%%%%%%%%%%%%%%%%%%%%%%%%%%
In order to make contact with a physical system one has to relate
the RMT parameters $\mu$ and $N$ with $\hbar$ and the parameters
of the physical system. This is done in the short time limit. This
limit is not universal and the dynamics of the chaotic system is
not described by RMT. It is used only to determine the relation
between the parameters. It will be assumed for concreteness that
the system we wish to model by a random matrix is a two
dimensional chaotic billiard (a free particle of mass $m$ in a two
dimensional box) of area ${\cal A}$. The results of the paper do
not depend on this assumption. First we establish a relation
between the mean density of states of this model to the one of
RMT. In the framework of RMT the semicircle law (\ref{semicircle})
can be used for the mean density of states. If the density in the
center of the strip of energies, modeled by the random matrix,
coincides with that of the two dimensional billiard,
\begin{equation}\label{constraint1}
\overline{\rho}_x(0)=\frac{2N}{\pi}=\sqrt{4\beta\mu^2N} \times
\overline{\rho}_{2d}(E)= \sqrt{4\beta\mu^2N} \times\frac{m {\cal A}}{2\pi
\hbar^2}.
\end{equation}
The existence of the semiclassical limit for the correlation
function (\ref{BRCijQ}) \cite{Feingold86} leads to another
constraint. This constraint enables the expression $\mu$ in terms
of $\hbar$. It turns out that in the semiclassical limit the
number of levels in a given interval grows with $N$. We shall give
the explicit connection between classical parameters of the
system, $\hbar$, $\mu$ and $N$ in what follows. For the model
(\ref{model}) and other models where statistical independence
between $dH(X)/dX$ and $H(X)$ holds, (\ref{CnNoCorr}) can be used.
In this framework it is easy to take the semiclassical limit. In
this limit the spectrum can be considered continuous for fixed
time, so that the sum can be replaced by a suitably weighted
integral:
\begin{equation}\label{CnApproxIntDef}
C_\beta(t)\equiv \tilde{C}_\beta(t) -\beta\mu^2
\end{equation}
where
\begin{equation}\label{CnApproxInt}
\tilde{C}_\beta(t)\approx\beta\mu^2 \int_0^\infty dx
\overline{\rho}_x(x)\cos\left[\frac{t}{\hbar}\sqrt{4\beta\mu^2N}
x\right] ~\mbox{for}~\hbar\rightarrow0,
\end{equation}
and $\beta\mu^2$ is the contribution of the $m=n$ term in
(\ref{CnNoCorr}). In the spirit of
the RMT modeling we choose the eigenvalue $E_n$ to be in the
middle of the region described by RMT, therefore we set $E_n=0$.
The integral in (\ref{CnApproxInt}) is known \cite{GR80}, and we
obtain:
\begin{equation}
\tilde{C}_\beta(t)\approx\sqrt{\beta\mu^2 N} \frac{\hbar}{t}
\func{J}_1\left[2\sqrt{\beta\mu^2 N} \frac{t}{\hbar}\right]
~\mbox{for}~\hbar\rightarrow0.
\end{equation}
The correlation function has a characteristic time scale
$T_c=\hbar/\sqrt{\beta\mu^2 N}$. Since the integral (\ref{InDef})
for $I_\beta(t)$ is convergent for all values of $t\gg T_c$, it is
well approximated by its value at $t\gtsimscript T_c$. Therefore,
$I_\beta(t=\infty)\equiv\overline{I}$ is expected to take a
classical value in the limit $N\rightarrow\infty$. One finds:
\begin{equation}\label{constraint2}
\overline{I} = \hbar \sqrt{\beta\mu^2 N}.
\end{equation}
Now units where ${\cal A}m/2\equiv2$ and $\overline{I}\equiv1/2$
are introduced. In such units $\mu$ and $\hbar$ are dimensionless
and are given in terms of $N$ as:
\begin{eqnarray}
\beta\mu^2 &=& \frac14 N^{-1/3},\\ \hbar &=& N^{-1/3}.
\end{eqnarray}
The Heisenberg time in these units is:
\begin{equation}\label{Heisenberg}
T_H\equiv h~\overline{\rho}_{2d}(E=0)=4N^{1/3},
\end{equation}
and
\begin{equation} \label{units}
\beta\mu^2T_H=1,
\end{equation}
while the characteristic time scale for the saturation of
$I_\beta$ to its classical value is $T_c=2 N^{-2/3}$. Finally, in
these units:
\begin{equation}\label{noname}
C_\beta(\tau)\approx
\frac{\func{J}_1\left[4N\tau\right]}{8N^{1/3}\tau}
-\frac{1}{4N^{1/3}} ~~~~\mbox{for}~N\rightarrow\infty,
\end{equation}
where $\tau\equiv t/T_H$ is the dimensionless time. The integral
over the correlation function is, in these units:
\begin{equation}\label{InLArgeN}
I_\beta(\tau)\approx 1/2 - \tau,
\end{equation}
and the approximation holds for $\tau\ll 1$. Now we can justify
some of the assumptions that we made. First of all, we see that
the limit $N\rightarrow\infty$ indeed corresponds to the limit
$\hbar\rightarrow0$. Secondly, the mean level spacing $\Delta
E=1/\overline{\rho}_{2d}(0)$ and $T_c$ decay to zero (as
$N^{-2/3}$) in the limit $N\rightarrow\infty$, as expected.

%%%%%%%%%%%%%%%%%%%%%%%%%%%%%%%%%%%%%%%%%%
\section{The long time behavior predicted from the nearest
neighbor level spacing distribution}\label{NND}
%%%%%%%%%%%%%%%%%%%%%%%%%%%%%%%%%%%%%%%%%
In the previous section the integral of the correlation function
$I_\beta(\tau)$ was studied for RMT models. One conclusion was
that it is dominated by the nearest neighbor level spacings. It
was found also that there is a big difference between GOE on the
one hand and GUE and GSE on the other. In this section we shall
study the decay of the correlation function if the nearest
neighbor level spacing distribution is given and will not rely on
the assumption of an invariant RMT ensemble. We will also assume
that the fluctuations of
$\left|\left(dH(X)/dX\right)_{n,m}\right|^2$ are not important and
this quantity can be replaced by the constant $\beta \mu^2$. It
will be assumed that the distribution of the nearest neighbor
level spacings is of the form:
\begin{equation}\label{Psbeta}
P_\beta(s)=c~s^\beta \exp{\left[-a s^2\right]}
\end{equation}
where $a$ and $c$ are constants. The RMT distributions
(\ref{PsGOE}), (\ref{PsGUE}) and (\ref{PsGSE}) for $\beta=1,2,4$
are of this form. The integral for the correlation function
(\ref{InApproxLong}) takes the form:
\begin{equation}\label{InCbeta}
I_\beta(\tau)/\beta \mu^2 T_H=\frac{c}{\pi} \int_0^\infty
\!\!\!\!ds~ s^{\beta-1} \exp{\left[-a s^2\right]} \sin(2 \pi \tau
s)=\frac{c}{\pi a ^{\beta/2}}{\cal I}_\beta (y)
\end{equation}
where
\begin{equation}\label{Jdef}
 {\cal I}_\beta(y)\equiv\int_0^\infty \!\!\!\!ds~ s^{\beta-1}
  \exp{[-s^2]}\sin{s y},
\end{equation}
with $y=2 \pi \tau /\sqrt{a}$. The decay of this function will be
explored in what follows, for arbitrary $\beta$. One can verify
that this function satisfies the ordinary differential equation:
\begin{equation}\label{DiffEqn}
  \left\{\begin{array}{l}
    \frac{d^2}{dy^2}{{\cal I}}_\beta(y)+\frac{y}{2}\frac{d}{dy}
  {{\cal I}}_\beta(y)+\frac{\beta}2
  {\cal I}_\beta(y)=0 \\
    {\cal I}_\beta(0)=0;~~
    \frac{d}{dy}{{\cal I}}_\beta(0)=\frac12\Gamma\left[\frac{\beta+1}2\right]. \
  \end{array}\right.
\end{equation}
Making the substitution:
\begin{equation}\label{DiffEqn2}
  {\cal I}_\beta(y)=f_\beta(y/\sqrt{2})\exp{\left[-y^2/8\right]}
\end{equation}
and changing to the variable $x=y/\sqrt{2}$, one arrives at a
new differential equation:
\begin{equation}\label{DiffEqn3}
\left\{  \begin{array}{l}
    f''(x)-\left[x^2/4 + 1/2 -\beta\right]f(x)=0\\
    f_\beta(0)=0;~~
    f'_\beta(0)=\frac12\Gamma\left[\frac{\beta+1}2\right]. \
  \end{array}\right.
\end{equation}
Eq.~\ref{DiffEqn3} is a well known equation, and its solutions are
Parabolic Cylinder Functions \cite{AS}: $\func{U}\left[1/2
-\beta,x\right], \func{V}\left[1/2 -\beta,x\right]$. For arbitrary
$\beta$, the solution of (\ref{DiffEqn3}) is a linear combination
of these two functions:
\begin{equation}\label{SolGen}
  f_\beta(x)=A_\beta~\func{U}\!\left[\frac12-\beta,x\right] +
  B_\beta~\func{V}\!\left[\frac12-\beta,x\right],
\end{equation}
in which $A_\beta, B_\beta$ are constants to be determined from
the initial conditions. In particular, if $\beta$ is an odd
natural number it turns out that $A_\beta=0$, if it is an even
natural number $B_\beta=0$ while for non-integer $\beta$ both
$A_\beta$ and $B_\beta$ are non-vanishing. This fact is very
important for the asymptotic behavior of ${\cal I}_\beta(y)$,
which is determined by the large $x$ behavior of $U\left[\frac12
-\beta,x\right], V\left[\frac12 -\beta,x\right]$ \cite{AS}:
\begin{eqnarray}
\label{Uasymp} \func{U}\!\left[\frac12
-\beta,x\right]&\sim&x^{\beta-1}\exp{\left[-x^2/4\right]}\\
\label{Vasymp} \func{V}\!\left[\frac12
-\beta,x\right]&\sim&\sqrt{\frac2\pi}~x^{-\beta}\exp{\left[x^2/4\right]},
\end{eqnarray}
as $x\rightarrow\infty$. One can immediately deduce that the first
solution is subdominant for all $\beta$ except for the special
case when it is an even natural number, for which $B_\beta=0$ in
(\ref{SolGen}). Since ${\cal I}_\beta (y)$ is proportional to
$I_\beta(\tau)$ and $y$ is proportional to $\tau$, for any fixed
$\beta\neq2n$ ($n=1,2,3,\ldots$):
\begin{equation}\label{JasympMost}
  I_\beta(\tau)\sim \tau^{-\beta},
\end{equation}
as $\tau\rightarrow\infty$, while for $\beta=2n$
($n=1,2,3,\ldots$):
\begin{equation}\label{JasympEven}
  I_\beta(\tau)\sim
  \exp{\left[-\frac{\pi^2}{a}\tau^2\right]}\tau^{\beta-1},
\end{equation}
as $\tau\rightarrow\infty$. This is precisely the type of behavior
found for the random matrix ensembles treated explicitly
(\ref{InApproxLong2GOE}, \ref{InApproxLong1GUE} \&
\ref{InApproxLongGSE}).

Finally, one wonders what would the analogous results be in the
case of the Poisson distribution. Returning to
(\ref{CnApproxLong0}), one obtains for the correlation function:
\begin{equation}\label{CnPoisson0}
C_P(\tau)\approx 2\sigma_P^2\int_0^\infty \!\!\!\!ds~ P_P(s)
\cos{(2 \pi\tau s)},
\end{equation}
where $\sigma_P^2$ is the variance of the off-diagonal matrix
elements between nearest neighbor levels, which we shall leave
unspecified, as we are only interested in the behavior as a
function of $\tau$. An approximation similar to the one leading to
(\ref{CnApproxLong1}) was made. For the Poisson case the nearest
neighbor spacing distribution is:
\begin{equation}\label{PsPoisson}
P_P(s)=\exp[-s],
\end{equation}
leading to:
\begin{equation}\label{CnPoisson1}
C_P(\tau)\approx\frac{2\sigma_P^2}{1+(2 \pi \tau)^2}.
\end{equation}
The integral over this expression is:
\begin{equation}\label{InPoisson1}
I_P(\tau)\approx \frac{\sigma_P^2 T_H}{\pi}\arctan{(2 \pi \tau)},
\end{equation}
the asymptotic behavior of which is given by:
\begin{equation}\label{InPoisson2}
I_P(\tau)\sim \frac{\sigma_P^2 T_H}{\pi}\left(\frac{\pi}2-\frac{1}{2 \pi \tau}+
\cdots\right) \rightarrow  \sigma_P^2 T_H/2,
\end{equation}
as $\tau\rightarrow\infty$. In the absence of level repulsion one
indeed finds that the integral of the correlation function does
not vanish. This does not result in any contradiction with the
classical limit where for an integrable system $I(t=\infty)=0$.
The reason is that for integrable systems the eigenfunctions of
neighboring energy levels typically have an exponentially small
overlap (in $1/\hbar$). This small overlap is the physical reason
for the Poisson distribution. Therefore in the classical limit
$\sigma_P^2 \rightarrow 0$ exponentially fast in $1/\hbar$.

Many systems that are neither integrable nor chaotic were found to
have a semi-Poisson distribution \cite{Bog99}. For this case, that
shows linear level repulsion, the nearest neighbor spacing
distribution is:
\begin{equation}\label{PsmPoisson}
P_{SP}(s)=4s\exp[-2s],
\end{equation}
The integral of the correlation function was calculated along the
lines of the calculation for the Poisson distribution. The result
is:
\begin{equation}\label{InSPoisson1}
I_{SP}(\tau)\approx 2 \sigma_{SP}^2 T_H \frac{\tau}{1+( \pi
\tau)^2}.
\end{equation}
It decays like $1/\tau$ in the long time limit. For the
semi-Poisson distribution there is level repulsion and indeed the
integral of the correlation function decays with time.

\begin{figure}
\centering\epsfig{file=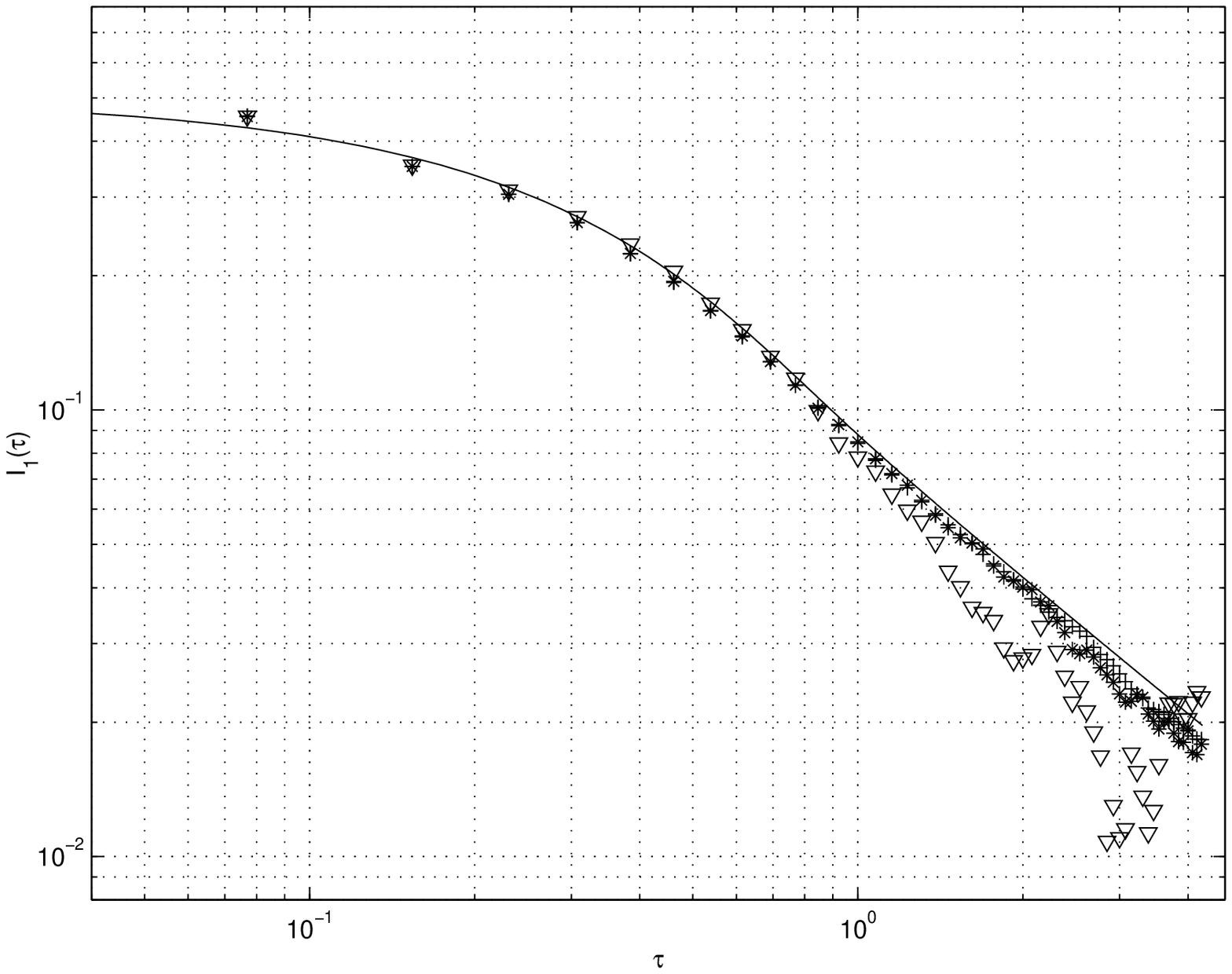,height=10cm,angle=90}
\caption{Testing the dependence on the size of the ensemble for
GOE with $N=13$ where all level spacings are taken into account:
$10^4$ members ($\scriptstyle\bigtriangledown$), $10^5$ members
($\ast$) and $10^6$ members ($\scriptstyle+$). The errors are of
order $\Delta I_1\approx0.01$, $\Delta I_1\approx0.005$ and
$\Delta I_1\approx0.001$ respectively. Also shown is the long $t$
approximation (Eq.~\protect\ref{InApproxLong1GOE})
(line)}\label{figGOEfluc}
\end{figure}

\begin{figure}
\centering\epsfig{file=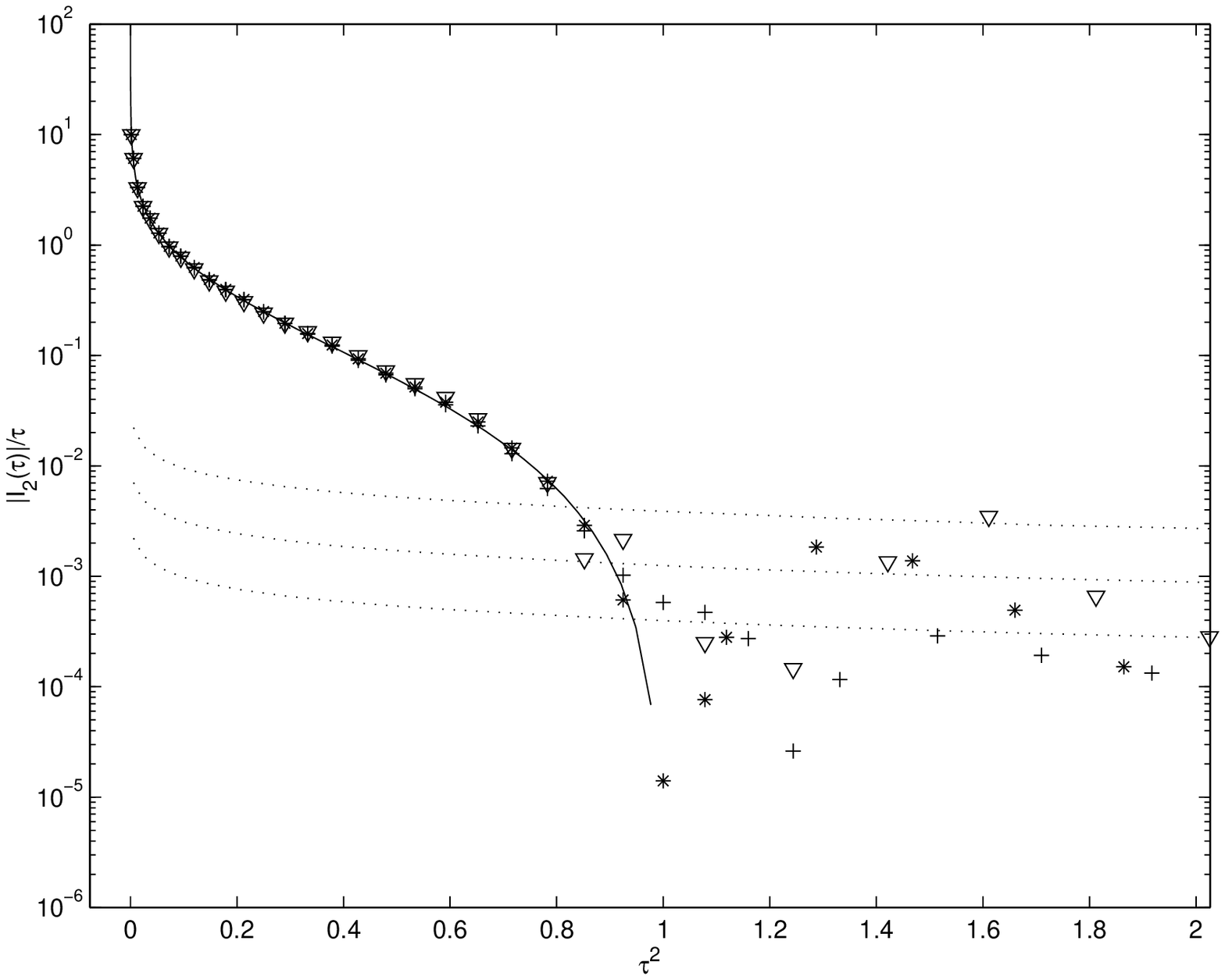,height=10cm,angle=90}
\caption{Testing the dependence on the size of the ensemble for
GUE with $N=13$ where all level spacings are taken into account:
$10^4$ members ($\scriptstyle\bigtriangledown$), $10^5$ members
($\ast$) and $10^6$ members ($\scriptstyle+$). The errors are of
order $\Delta I_2\approx10^{-3}$, $\Delta I_2\approx10^{-4}$ and
$\Delta I_2\approx10^{-5}$ respectively. Also shown is the long
$t$ approximation (Eq.~\protect\ref{InApproxLong1GUE}) (solid
line). The errors are marked by dotted lines, the top line is for
$10^4$, the middle for $10^5$ and the bottom for $10^6$.}
\label{figGUEfluc}
\end{figure}

%%%%%%%%%%%%%%%%%%%%%%%
\section{Summary and Discussion}\label{SecDiscussion}
%%%%%%%%%%%%%%%%%%%%%%%
The correlation function of the force applied by a fast quantum
system on a slow classical one is calculated within the leading
order correction to the adiabatic approximation following
Berry-Robbins and Jarzynski. Its finite time integral $I(t)$ of
(\ref{InDef}) is proportional to the dissipation rate on the time
scale $t$. In the present work $I(t)$ was calculated under various
statistical assumptions. In Section \ref{SecRMT} it was studied in
the framework of RMT. For the specific dependence of the model
(\ref{model}) on the parameter $X$, the Hamiltonian $H$ and
$dH/dX$ are statistically independent \cite{Austin92}. For this
case it was shown that up to a proportionality constant, the
integral of the correlation function is simply related to the
integral of the form factor (by Eq.~\ref{IandK}). Since the form
factor is known in RMT, the integral of the correlation function
was calculated and found to vanish for times beyond the Heisenberg
time for GUE and to fall off as a power law for GOE. This is a
remarkable and surprising difference. The result for GSE is
similar to the one found for GUE, except that the integral of the
correlation function vanishes after twice the Heisenberg time. The
properties of the model (\ref{model}) are satisfied approximately
by many systems \cite{lerner}, therefore it is expected that the
results of this work are relevant for a wide range of problems.
For the model (\ref{model}) we have shown that for long times the
results are dominated by the nearest neighbor spacings. If only
these spacings are taken into account one finds that for long
times the integral falls off as a power law for GOE and as a
Gaussian, where the characteristic time is proportional to the
Heisenberg time for GUE and GSE. These results do not require all
the properties of the model (\ref{model}). They require only that
the contribution of the fluctuations in the absolute value of the
matrix elements between nearest neighboring states (neighboring in
energy) is negligible, namely that the contribution of $\Delta
C_\beta(t)$ of (\ref{DCn}) can be ignored compared to the one of
$C_\beta(t)$. This is clearly a weaker assumption than complete
statistical independence between $H$ and $dH/dX$. Therefore we
expect the difference found between GOE and GUE (as well as GSE)
to be generic for RMT models with various dependencies on the
parameter $X$. The long time behavior of the RMT models is
expected to provide a faithful representation of the behavior of
chaotic systems, since it is dominated by the small level
spacings. For short times, on the other hand, the behavior depends
on the specific properties of each system. The short time behavior
of the RMT model was presented here only to set the relation
between the constants of the RMT model and the ones of the chaotic
system.

The assumption of the dominance of the contribution of nearest
neighbor level spacings, together with the assumption that $\Delta
C_\beta(t)$ of (\ref{DCn}) is negligible enables the calculation
of the integral $I(t)$ of (\ref{InDef}) for various distributions
of nearest neighbor level spacings even if these do not
necessarily originate from RMT models. For the distribution
(\ref{Psbeta}), that is a generalization of the distributions
found for GOE, GUE and GSE, one can calculate $I_\beta(t)$ for
various values of $\beta$. In Section \ref{NND} it is found to
decay as $t^{-\beta}$ for all values of $\beta$, except when
$\beta$ is a positive even integer, for which it decays like a
Gaussian with a characteristic time that is proportional to the
Heisenberg time. What is special when $\beta$ is a positive even
integer? For these values the integral (\ref{InCbeta}) can be
extended to the range $[-\infty, \infty]$. The integrand is an
entire function, the contour of integration can be deformed in the
complex plane and the integral is dominated by a saddle point. For
other values of $\beta$ an extension of the integral to negative
$t$, so that the integrand is analytic, is impossible. The point
$s=0$ is an end point and for large $t$ the integral is dominated
by it, leading to power law decay. For non-integral $\beta$, the
point $s=0$ is also a singular point. It would be nice to find a
more physical explanation for this difference between the various
ensembles. For completeness the integral of the correlation
function was calculated for the Poisson and the semi-Poisson
distributions.

The various RMT formulas (Eqs.~\ref{IGOE}~\&~\ref{IGUE}) are
developed for the limit of infinite matrices. This limit is
approached extremely fast, as can be seen in
Figs.~\ref{figGOE}~\&~\ref{figGUE}. The convergence to the
average, as a function of $N_{ens}$, the number of members of the
ensemble, is slow (see
Figs.~\ref{figGOEfluc}~\&~\ref{figGUEfluc}).

The crucial approximation that was made generalizing the results
beyond the model (\ref{model}) was neglecting $\Delta C_\beta(t)$
of (\ref{DCn}). Although reasonable, its validity for chaotic
systems should be checked. The results may hold also for mixed
systems if sticking to integrable regions does not take place on
time scales relevant for the calculation. For chaotic systems
corrections of order higher than the leading one, in the adiabatic
approximation, may lead to different behavior after some time
($T_2$ or $T_{LZ}$). Dephasing, as a result of the coupling to the
environment will destroy the quantum correlations on a time scale
$T_\phi$. In experiments of the type mentioned in the end of the
Introduction, dephasing is always present, and if $T_\phi\ll T_2,
T_{LZ}$, the system will dissipate energy at a rate proportional
to $I_\beta(T_\phi)$.

%%%%%%%%%%%%%%%%%%%%%%%
\section{Acknowledgments}
%%%%%%%%%%%%%%%%%%%%%%%
We have benefited from discussions with M. Berry, E. Bogomolny, D.
Cohen, B. Eckhardt, J. Feinberg, F. Izrailev, C. Jarzynski, C.
Marcus, E. Ott, R. Prange, J. Robbins, H.-J. St\"{o}ckmann, M.
Wilkinson and M. Zirnbauer. We thank in particular B. Simons for
extremely illuminating remarks. This research was supported in
part by the U.S.--Israel Binational Science Foundation (BSF), by
the Minerva Center for Non-linear Physics of Complex Systems, by
the Israel Science Foundation, by the Niedersachsen Ministry of
Science (Germany) and by the Fund for Promotion of Research at the
Technion. One of us (SF) would like to thank R. Prange for the
hospitality at the University of Maryland where this work was
completed.


\begin{thebibliography}{10}

\bibitem{Hill52}
D.~L. Hill and J.~A. Wheeler.
\newblock {\em Phys. Rev}, 89:1102, 1952.

\bibitem{Blocki78}
J.~Blocki, Y.~Boneh, J.~R. Nix, J.~Randrup, M.~Robel, A.~J. Sierk,
and W.~J. Swiatecki.
\newblock {\em Ann. of Phys.}, 113:330, 1978;
S.~E. Koonin, R.~L. Hatch, and J.~Randrup.
\newblock {\em Nucl. Phys.}, A283:87, 1977;
S.~E. Koonin and J.~Randrup.
\newblock {\em Nucl. Phys.}, A289:475, 1977.

\bibitem{Burgio95}
G.~F. Burgio, M.~Baldo, A.~Rapisarda, and P.~Schuck.
\newblock {\em Phys. Rev. C}, 52:2475, 1995.

\bibitem{Wilkinson95}
M.~Wilkinson and E.~J. Austin.
\newblock {\em J. Phys. A: Math. Gen}, 28:2277, 1995;
M.~Wilkinson.
\newblock {\em J. Phys. A: Math. Gen.}, 21:4021, 1988.

\bibitem{Wilkinson90}
M.~Wilkinson.
\newblock {\em J. Phys. A: Math. Gen.}, 23:3603, 1990;
\newblock {\em J. Phys. A: Math. Gen.}, 20:2145,
1987.

\bibitem{Austin92}
E.~J. Austin and M.~Wilkinson.
\newblock {\em Nonlinearity}, 5:1137, 1992.

\bibitem{aberg} S.~Mizutori and S.~Aberg,
\newblock {\em Phys. Rev. E}, 56:6311, 1997.

\bibitem{Bulgac98}
A.~Bulgac, Gd. Dang, and D.~Kusnezov. \newblock {\em Phys. Rev.
E}, 58:196, 1998; \newblock {\em Phys. Rev. E}, 54:3468, 1996;
\newblock {\em Ann. Phys.}, 242:1, 1995; D.~Mitchell, Y.~Alhassid,
and D.~Kusnezov. \newblock {\em Phys. Lett. A}, 215:21, 1996.

\bibitem{CohenD98}
D.~Cohen.
\newblock {\em Phys. Rev. Lett}, 82:4951, 1999;
\newblock Lecture notes, to be published in
proceedings of Session CXLIII of the Enrico Fermi Summer-School on
``New Directions in Quantum Chaos'', Varenna 1999; \newblock
"Chaos and Energy Spreading for Time-Dependent Hamiltonians, and
the various Regimes in the Theory of Quantum Dissipation",
cond-mat/9902168.

\bibitem{Ott79}
E.~Ott.
\newblock {\em Phys. Rev. Lett}, 42:1628, 1979.

\bibitem{Brown87a}
R.~Brown, E.~Ott, and C.~Grebogi.
\newblock {\em J. Stat. Phys.}, 49:511, 1987;
\newblock {\em Phys. Rev. Lett.}, 59:1173, 1987.

\bibitem{Berry93}
M.~V. Berry and J.~M. Robbins.
\newblock {\em Proc. R. Soc. Lond. A}, 442:659, 1993.

\bibitem{Jarzynski95}
C.~Jarzynski.
\newblock {\em Phys. Rev. Lett.}, 74(15):2937, 1995.

\bibitem{Jarzynski93b}
C.~Jarzynski.
\newblock {\em Phys. Rev. Lett.}, 71(6):839, 1993.

\bibitem{Berry97a}
M.~V. Berry and E.~C. Sinclair.
\newblock {\em J. Phys. A: Math. Gen}, 30:2853, 1997.

\bibitem{Jarzynski92}
C.~Jarzynski.
\newblock {\em Phys. Rev. A}, 46:7498, 1992;
\newblock {\em Phys. Rev. E}, 48:4340, 1993.

\bibitem{Bohigas84}
O.~Bohigas, M.~J. Giannoni, and C.~Schmidt.
\newblock {\em Phys. Rev. Lett.}, 52:1, 1984;
O. Bohigas,  in {\em Chaos and Quantum Physics , {\em Proc. of
Les-Houches
  Summer School, Session LII, 1989}}, edited by M.~J. Giannoni, A. Voros, and
  J. Zinn-Justin (Elsevier, 1991).

\bibitem{izrailev}
G.~P. Berman, F.~M. Izrailev and O.~F. Smokotina,
\newblock {\em Phys. Lett. A}, 161:483, 1992.

\bibitem{Wigner56}
E.~P. Wigner,  in {\em Conference on Neutron Physics by
Time-of-Flight}, Oak
  Ridge Natl. Lab. Rep. {\bf ORNL-2309} (Gatlinburg, Tennessee,
  1956), p.\ 59.
Reproduced in \cite{Porter65}.


\bibitem{Landau55}
L. Landau and Y. Smorodinsky, {\em Lectures on the Theory of the
Atomic
  Nucleus} (State Tech.-Theoret. Lit. Press, Moscow, 1955), p.\ 93,
  translation, pp. 54-55. Plenum Press (Consultants Bureau), New York, 1958.
   Reproduced in \cite{Porter65}.

\bibitem{Porter65}
C.~E. Porter, {\em Statistical Theories of Spectra: Fluctuations} (Academic
  Press, New York, 1965). Equations cited in the paper are from
  the Introduction.

\bibitem{Mehta91}
M.~L. Mehta, {\em Random Matrices} (Academic Press, New York, 1991
- second edition).

\bibitem{DCPC} D.~Cohen, Private Communication.

\bibitem{marcus}
M. Switkes, C.~M. Marcus, K. Campman and A.~C. Gossard, {\em
Science} 283:1905, 1999 and references therein.

\bibitem{Andreev95}
A.~V. Andreev and B.~L. Altshuler.
\newblock {\em Phys. Rev. Lett.}, 75:902, 1995;
O.~Agam, B.~L. Altshuler, and A.~V. Andreev.
\newblock {\em Phys. Rev. Lett.}, 75:4389, 1995;
A.~V. Andreev, O.~Agam, and B.~L. Altshuler.
\newblock {\em Phys. Rev. Lett.}, 76:3947, 1996;
A.~V. Andreev, O.~Agam, B.~D. Simons, and B.~L. Altshuler.
\newblock {\em Nucl. Phys. B}, 482:536, 1996;
\newblock {\em Phys. Rev. Lett.}, 79:1778, 1997.
\bibitem{Bogomolny96}
E.~Bogomolny and J.~Keating.
\newblock {\em Phys. Rev. Lett}, 77:1472, 1996.

\bibitem{Zirnbauer99}
M.~R. Zirnbauer,  in {\em Suppersymmetry and Trace Formulae, Chaos
and
  Disorder}, edited by I.~V. Lerner, J.~P. Keating, and D.~E. Khmelnitskii
  (Kluwer Academic / Plenum Publishers, New York, 1999), p.\ 193.

\bibitem{Feingold90}
 M.~Feingold, R.~G. Littlejohn, S.~B. Solina, and J.~S.
Pehing.
\newblock {\em Phys. Lett. A}, 146:199, 1990;
M.~Feingold, D.~M. Leitner, and M.~Wilkinson.
\newblock {\em Phys. Rev. Lett.}, 66:986, 1991;
M.~Wilkinson, M.~Feingold, and D.~M. Leitner.
\newblock {\em J. Phys. A: Math. Gen.}, 24:175, 1991.

\bibitem{Wigner55}
E.~P. Wigner.
\newblock {\em Ann. Math.}, 62:548, 1955. Reproduced in \cite{Porter65}.

\bibitem{lerner}
I.~T. Chalker, I.~V. Lerner and R.~A. Smith,
\newblock {\em Phys. Rev. Lett.}, 77:554, 1996;
\newblock {\em J. Math. Phys.}, 37:5061, 1996.

\bibitem{AS}
M. Abramowitz and A. Stegun, {\em Handbook of Mathematical
Functions} (National
  Bureau of Standards, Washington, D.C., 1964).

\bibitem{GR80}
I.~S. Gradshteyn and I.~M. Ryzhik, {\em Table of Integrals,
Series, and
  Products} (Academic Press, U.S.A., 1980).

\bibitem{Feingold86}
M.~Feingold and A.~Peres.
\newblock {\em Phys. Rev. A}, 34:591, 1986;
M.~Wilkinson.
\newblock {\em J. Phys. A: Math. Gen}, 20:2415, 1987.

\bibitem{Bog99}
E.~B. Bogomolny, U.~Gerland and C.~Schmit,
\newblock {\em Phys. Rev. E}, 59:R1315, 1999.

\end{thebibliography}
\end{document}